\documentclass[fleqn,usenatbib,useAMS]{mnras}

\usepackage[T1]{fontenc}
\usepackage{ae,aecompl}

\usepackage{graphicx}	
\usepackage{amsmath}	
\usepackage{amssymb}	

\usepackage[normalem]{ulem}

\newcommand{\kms}{{\,km\,s$^{-1}$}}

\title[The Vela supercluster] {Discovery of a supercluster in the Zone of Avoidance in Vela}

\author[Ren\'ee C. Kraan-Korteweg et al.]{Ren\'ee C. Kraan-Korteweg,$^{1}$\thanks{E-mail: \href{mailto:kraan@ast.uct.ac.za}{kraan@ast.uct.ac.za}}
Michelle E. Cluver,$^{2}$\thanks{E-mail: \href{mailto:michelle.cluver@gmail.com}{michelle.cluver@gmail.com}}
Maciej Bilicki,$^{3,4}$\thanks{E-mail: \href{mailto:bilicki@strw.leidenuniv.nl}{bilicki@strw.leidenuniv.nl}}
\newauthor
Thomas H. Jarrett,$^{1}$
Matthew Colless,$^{5}$
Ahmed Elagali,$^{1}$
Hans B\"ohringer$^{6}$
\newauthor
and 
Gayoung Chon$^{6}$
\\
$^{1}$Department of Astronomy, University of Cape Town, 7700 Rondebosch, South Africa\\
$^{2}$Department of Physics and Astronomy, University of the Western Cape, 7535 Bellville, South Africa\\
$^{3}$Leiden Observatory, Leiden University, Niels Bohrweg 2, NL-2333 CA Leiden, The Netherlands \\ 
$^{4}$Janusz Gil Institute of Astronomy, University of Zielona G\'ora, ul.\ Licealna 9, 65-417 Zielona G\'{o}ra, Poland \\
$^{5}$Research School of Astronomy and Astrophysics, Australian National University, Canberra, ACT 2611, Australia\\
$^{6}$Max-Planck-Institut f\"ur extraterrestrische Physik, D-85748 Garching, Germany}

\date{Accepted to MNRAS Letters, on 04 November 2016}

\pubyear{2016}

\begin{document}
\label{firstpage}
\pagerange{\pageref{firstpage}--\pageref{lastpage}}
\maketitle

\begin{abstract}
We report the discovery of a potentially major supercluster that extends across the Galactic Plane in the constellation of Vela, at a mean recessional velocity of $\sim$18\,000\kms. Recent multi-object spectroscopic observations of this Vela Supercluster (VSCL), using AAOmega+2dF and the Southern African Large Telescope, confirm an extended galaxy overdensity in the Zone of Avoidance (ZOA) located where residual bulk flows predict a considerable mass excess. We present a preliminary analysis of $\sim$4500 new spectroscopic galaxy redshifts obtained in the ZOA centred on the Vela region ($l = 272\fdg5 \pm 20\degr, b = 0\degr \pm 10\degr$). The presently sparsely-sampled dataset traces an overdensity that covers $25\degr$ in Galactic longitude on either side of the Plane, suggesting an extent of $25\degr \times 20\degr$, corresponding to $\sim$\,115$\,\times\,$90~$h_{70}$~Mpc at the supercluster redshift. In redshift space, the overdensity appears to consist of two merging wall-like structures, interspersed with clusters and groups. Both the velocity histogram and the morphology of the multi-branching wall structure are consistent with a supercluster classification. $K_s^o$ galaxy counts show an enhancement of $\sim1.2$ over the survey area for galaxies brighter than $M_{\rm K}^*$ at the VSCL distance, and a galaxy overdensity of $\delta=0.50$--$0.77$ within a photometric redshift shell around the VSCL, when compared to various Two-Micron All-Sky Survey samples. Taking account of selection effects, the VSCL is estimated to contribute $v_\mathrm{LG} \ga 50$ \kms\ to the motion of the Local Group.
\end{abstract}




\section{Introduction}


Galaxies are distributed in a cosmic web consisting of high-density clusters and galaxy groups embedded in walls and filaments surrounding large low-density regions \citep[e.g.][]{Jones09, SDSS}. The largest agglomerations, superclusters, can contain dozens of massive galaxy clusters \citep{Proust06, Chon13, Einasto14}. Such galaxy overdensities exert gravitational perturbations on the smoothly expanding Universe and induce coherent flows over large volumes, as, for instance, the Local Group (LG) motion imprinted as a dipole in the cosmic microwave background  ($\sim622$\kms; \citealt{Fixsen96}).

Despite numerous studies \citep[e.g.][]{Hudson04, Erdogdu06,  Kocevski06, Watkins09, Lavaux10, Bilicki11, Branchini12, CosFlow2, Springob16}, the resulting direction and amplitude of the LG peculiar velocity and of local bulk flows remain controversial. The persistent discrepancies may originate, in large part, from the incomplete mapping of large-scale structures in the Zone of Avoidance \citep[ZOA;][]{ZOAR00, Loeb08} which are excluded in `whole-sky' galaxy surveys, {\sl and} observed with lower sensitivity at the relevant higher distance range ($cz \gtrsim 16\,000$\kms) in targeted surveys

To reconcile the discrepancies, a hidden mass overdensity was postulated behind the southern ZOA \citep{Loeb08}. One possibility, a supercluster at or beyond the Shapley Concentration (SSC), with $cz > 16\,000$\kms, was considered unlikely even if it resided in the ZOA, as there had been no indication of its existence in whole-sky surveys. Since then, however, various new studies implied a considerable mass excess close to the ZOA in the direction of the Vela constellation at higher redshifts \citep[e.g.][]{Hudson04, Feldman10, Nusser11, Carrick15, CosFlow2}. And the recent 6dFGS and 2MTF peculiar velocity analyses \citep{Springob14, Scrimgeour16} require a residual bulk flow (273\kms) arising from that same general direction, which is generated beyond a distance of $230~h_{70}$~Mpc. However, at these depths, the Vela area is not mapped by current spectroscopic surveys, which are too shallow or do not sample $|b|<10^\circ$ (e.g.\ 2MASS Redshift Survey, \citealt{Huchra12}, or 2M++, \citealt{LH11}). 

We have embarked on a long-term programme to map the large-scale galaxy distribution behind the Milky Way, giving particular emphasis to the Vela ZOA area ($l = 272\fdg5 \pm 20\degr, b = 0\degr \pm 10\degr$). Our earliest ZOA multi-object spectroscopic data close to Vela revealed clear hints of a substantial galaxy overdensity at 18\,000\kms\ \citep[see Figs. 3 \& 5 in][]{KK94} based on observations made with Optopus on the ESO 3.6m telescope in the Hydra/Antlia ZOA galaxy survey \citep[$295\degr \gtrsim l \gtrsim 275\degr$;][]{KK00}. Speculations about a possible connection across the ZOA between the Horologium-Reticulum and Shapley superclusters were alluded to. Subsequent observations of galaxies in the adjacent Vela ZOA region \citep[$275\degr \gtrsim l \gtrsim 250\degr$;][]{ZOAR00} with the 6dF multi-fibre spectrograph, on the UK Schmidt Telescope, of two observed contiguous ZOA fields (350 redshifts) revealed a highly significant peak at the same redshift. The combined dataset implies a galaxy overdensity of surprising extent for its mean redshift. An analysis of the extinction-corrected magnitudes of galaxies in the deep optical Vela ZOA catalogue suggested further clustering at this approximate distance range, and also on the other side of the Galactic Plane, linking up to two X-ray clusters \citep[CIZA J0812.5-5714 and CIZA J0820.9-5704;][]{Kocevski06} at a similar redshift ($\sim$18\,500\kms).

A wide-area observing campaign was therefore launched to assess the full extent of the Vela overdensity, using multi-object spectrographs on both the Southern African Large Telescope (SALT) and the 3.9m Anglo-Australian Telescope (AAT). Section~\ref{sect:data} will summarize the spectroscopic survey results, inclusive of the above-mentioned unpublished Optopus and 6dF redshifts, and Section~\ref{sect:space} describes the resulting velocity distribution. Section~\ref{sect:discuss} presents a discussion in support of a supercluster morphology and an assessment of the amplitude of the overdensity. The conclusions of this Letter and planned follow-up are discussed in Section~\ref{sect:conc}.

\section{Spectroscopic Data}\label{sect:data}

The spectroscopic survey focused on the ZOA region around Vela: $245\degr \lesssim l \lesssim 295\degr$ and $|b| \lesssim 10\degr$. The galaxy target sample was based on a combination of galaxies from  deep optical galaxy surveys \citep{ZOAR00, KK00}, which had the specific aim of reducing the unmapped area in the southern Galactic ZOA, plus galaxies identified in the 2MASS Extended Source Catalogue (2MASX) \citep{2MASX}. The optical catalogue is a diameter-selected sample ($D \geq 12\arcsec$) which was found by \cite{KK00} to be complete to a magnitude limit $B_J \leq 18\fm5 - 19\fm0$ up to extinction levels $A_{\rm B} = 3\fm0$. The 2MASX finds galaxies closer to the Galactic plane (up to about $A_{\rm B} \lesssim 10^{\rm m}$) but becomes less effective where the star density is high, and has a completeness limit of $K_{s} \lesssim 13\fm5$. The catalogues are complementary in the types of galaxies they uncover at low Galactic latitudes \citep{KKTJ05}.

From 2012 to 2014, when the SALT multi-object spectrograph (MOS) became operational, time was allocated to observe Vela fields of particularly high galaxy concentrations (undersampled by 6dF observations). With its numerous slitlets over the relatively small field of view ($8\arcmin \times 8\arcmin$), the SALT MOS is ideal for observing cluster cores in this distance range. With typical exposure times of 12--20~min, reliable redshifts ($\sigma_z \sim 150$\kms) were obtained for about 80\% of the potential cluster galaxies. The SALT observations identified 10 clusters (out of 13 targeted fields). This includes the two above-mentioned CIZA X-ray clusters, which had only one or two prior redshifts. Interestingly, 7 of the 10 clusters were found to reside in the redshift range of the Vela overdensity.

\begin{figure}
	\includegraphics[width=0.95\columnwidth]{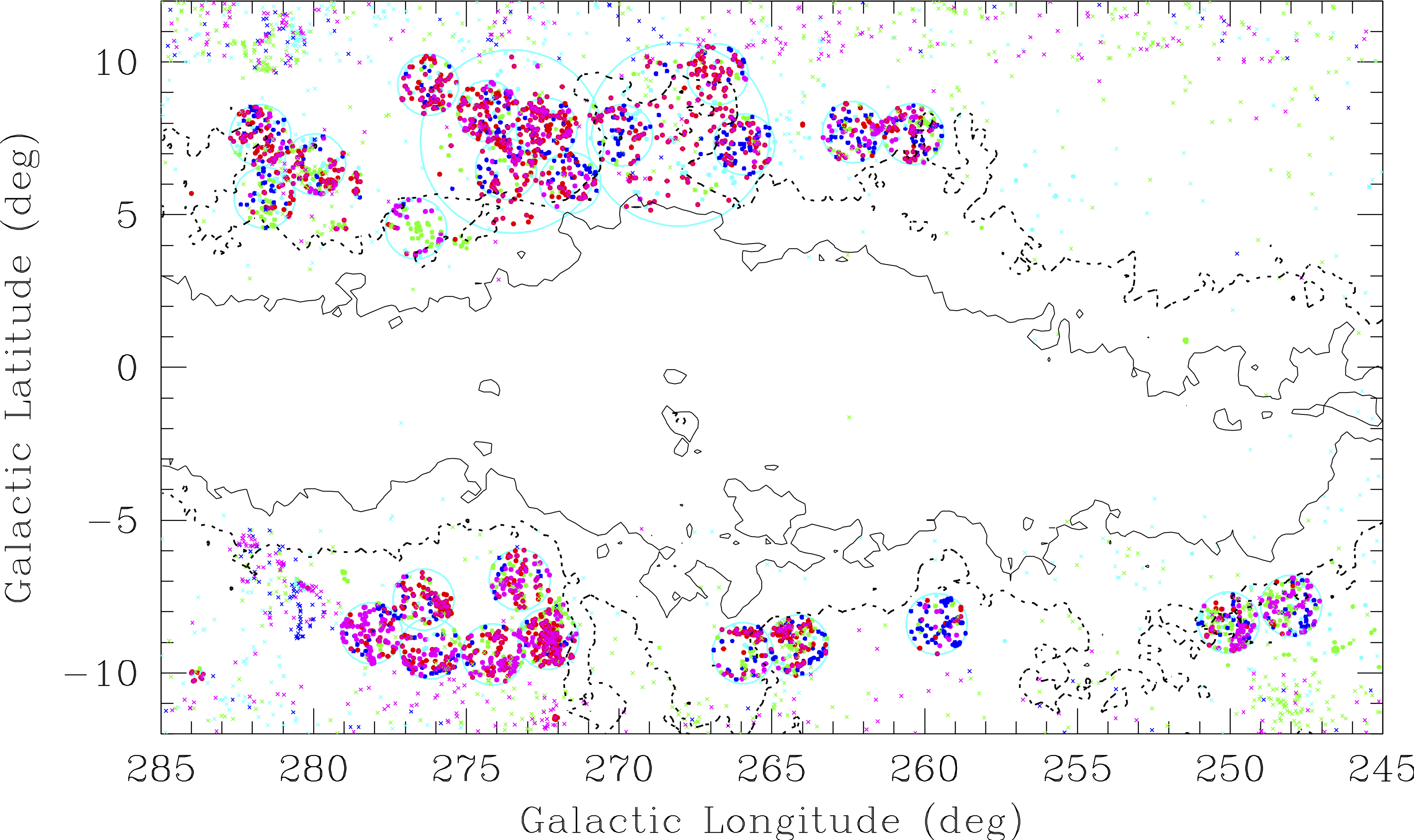}
    \caption{Redshift distribution in Galactic coordinates. Large dots mark new redshifts ($N=4432$), and crosses are previous redshifts. Colours indicate redshift (cyan: $<$8000; green: 8--16000; red: 16--20000; magenta: 20--24000; blue: 24--32000\kms). Cyan circles mark the two 6dF fields (large) and the 25 2\degr\ AAOmega fields (small). Contours indicate extinctions of $A_K = 0\fm3$ (solid), above which the galaxy sample is incomplete, and $0\fm18$ (dotted).}
    \label{fig:velonsky}
\end{figure}

Major strides were achieved in 2014 using the 2dF$+$AAOmega spectrograph on the 3.9m AAT, which has a 2\degr\ diameter field, in which up to 392 fibres can be positioned. We observed 25 fields with AAOmega over six nights. We employed a sparse-sampling strategy to cover as wide an area as possible in the allocated observing period; the locations of the fields are shown in Fig.~\ref{fig:velonsky}. Most of the pointings are towards intermediate latitudes ($|b| \sim 4\degr$--10.5\degr), where extinction is not excessive. Of the 4747 target galaxies in these 25 fields, 92\% could be fibred up. Exposure times were 60--90~min depending on foreground extinction and weather conditions. The success rate for science quality redshifts ($\sigma_z\sim 100$\kms) of these partially obscured galaxies was 95\%. These AAOmega observations yielded over 4100 unique new redshifts in the ZOA. 

Combining the AAOmega and SALT redshift data with our earlier unpublished data (Optopus, 6dF, and some 1.9m spectroscopy from SAAO), a total of 4432 new redshifts were obtained (although 126 lie at $|b| > 10\degr$). These galaxies and their redshifts are plotted in Galactic coordinates in Fig.~\ref{fig:velonsky}, including additional galaxy redshifts available from the literature. Note the scarcity of redshifts at $|b| \le 10\degr$, except at higher longitudes ($l \gtrsim 280\degr$) which our team surveyed in earlier work \citep{KK94}. The cluster cores targeted with SALT (too small and dense to be outlined separately in the Figure) mostly lie within the AAOmega fields, apart from one field at an extremely low latitude, $(l,b) \sim (251\fdg5,0\fdg9)$ and two CIZA X-ray clusters (at $b\sim-11\fdg5$ and $-12\fdg5$).

Despite the sparse sampling within the latitude bands of intermediate extinction ($|b| \gtrsim 4\degr$), the predominance of galaxies in the redshift range of the Vela overdensity (red and magenta; 16--24\,000\kms) is prominent in 20 of the 25 widely-spread AAOmega fields on both sides of the Galactic Plane. The overdensity appears to extend over most of our ZOA survey region, from about 25$\degr$ above to $20\degr$ below the Galactic Plane. 

\section{Distribution in redshift space}\label{sect:space}

\begin{figure}
\begin{center}	\includegraphics[width=0.88\columnwidth]{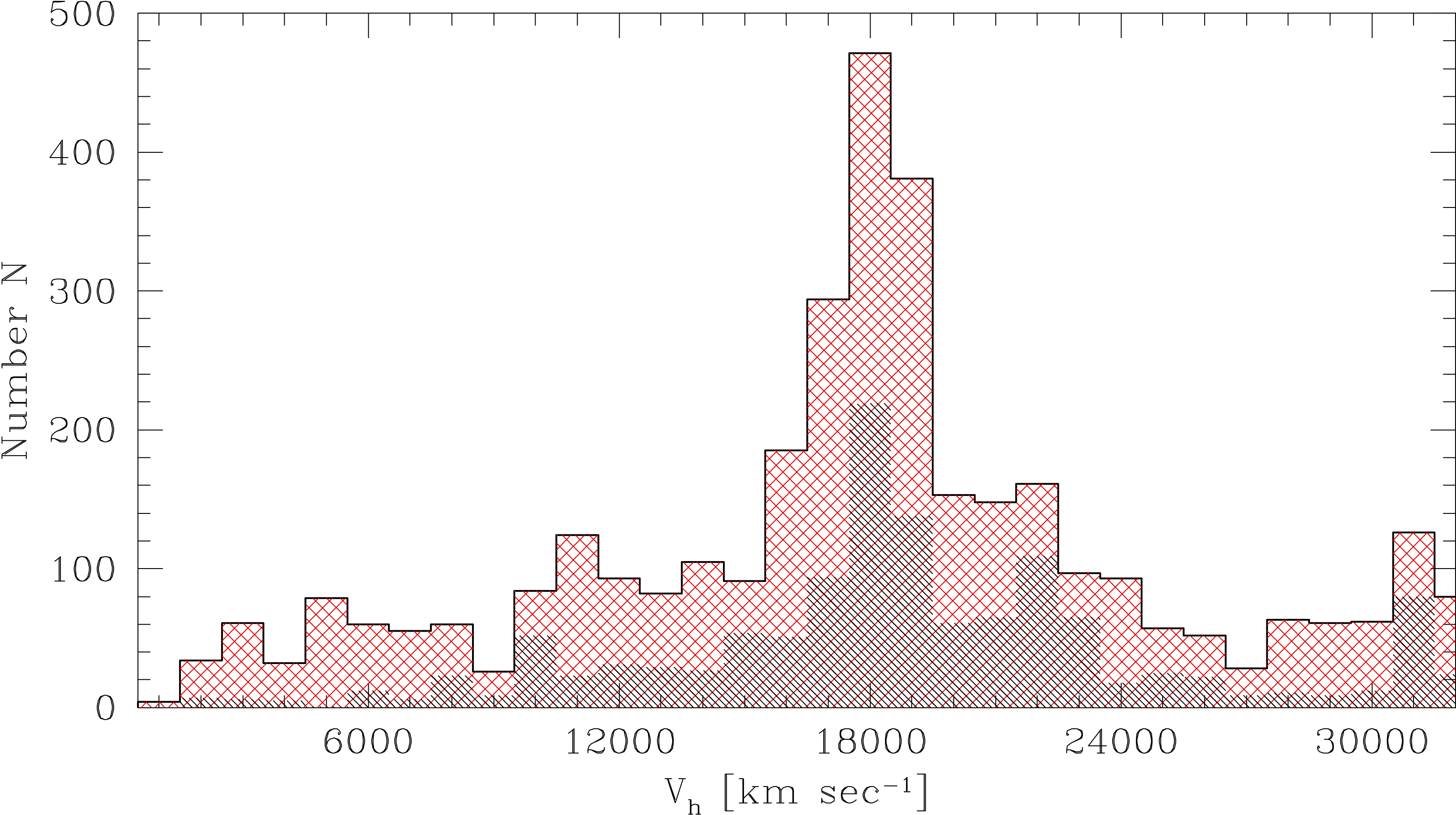}
    \caption{Redshift distribution in the Vela region ($255\degr \le l \le 285\degr, |b| \le 10\degr)$. Red shows all galaxies; grey shows galaxies below the Galactic Plane.} 
    \label{fig:vhist}
\end{center}
\end{figure}

To assess the significance of the overdensity, Fig.~\ref{fig:vhist} shows the redshift distribution of all galaxies in the region $255\degr \le l \le 285\degr$ and $|b| < 10\degr$. There is a highly significant peak at 17--19\,000\kms, with broad shoulders extending over 15--23\,000\kms. Such an extent is typical of superclusters \citep{Einasto14}, and, indeed, the velocity distribution is remarkably similar to that of the Shapley Supercluster (SSC) obtained by \cite{Proust06}. We note that the Vela structure emerges at a distance $cz \sim 16\,000$\kms\ where existing `whole-sky' redshift surveys lose sensitivity \citep{Jones09,Huchra12,CF3}.

The grey-shaded area marks the redshift distribution of galaxies below the Galactic Plane to emphasize that the overdensity is equally pronounced on either side of the Plane, with the galaxies below the Plane showing a secondary peak at $cz \sim 22\,000$\kms. The similarity of the histogram suggests that the overdensity is contiguous across the Plane, implying an overall extent of at least $25\degr \times 20\degr$, corresponding to $\sim$\,115$\,\times\,$90~$h_{70}$~Mpc.

\begin{figure*}
\begin{center}
\includegraphics[scale=0.55,angle=0]{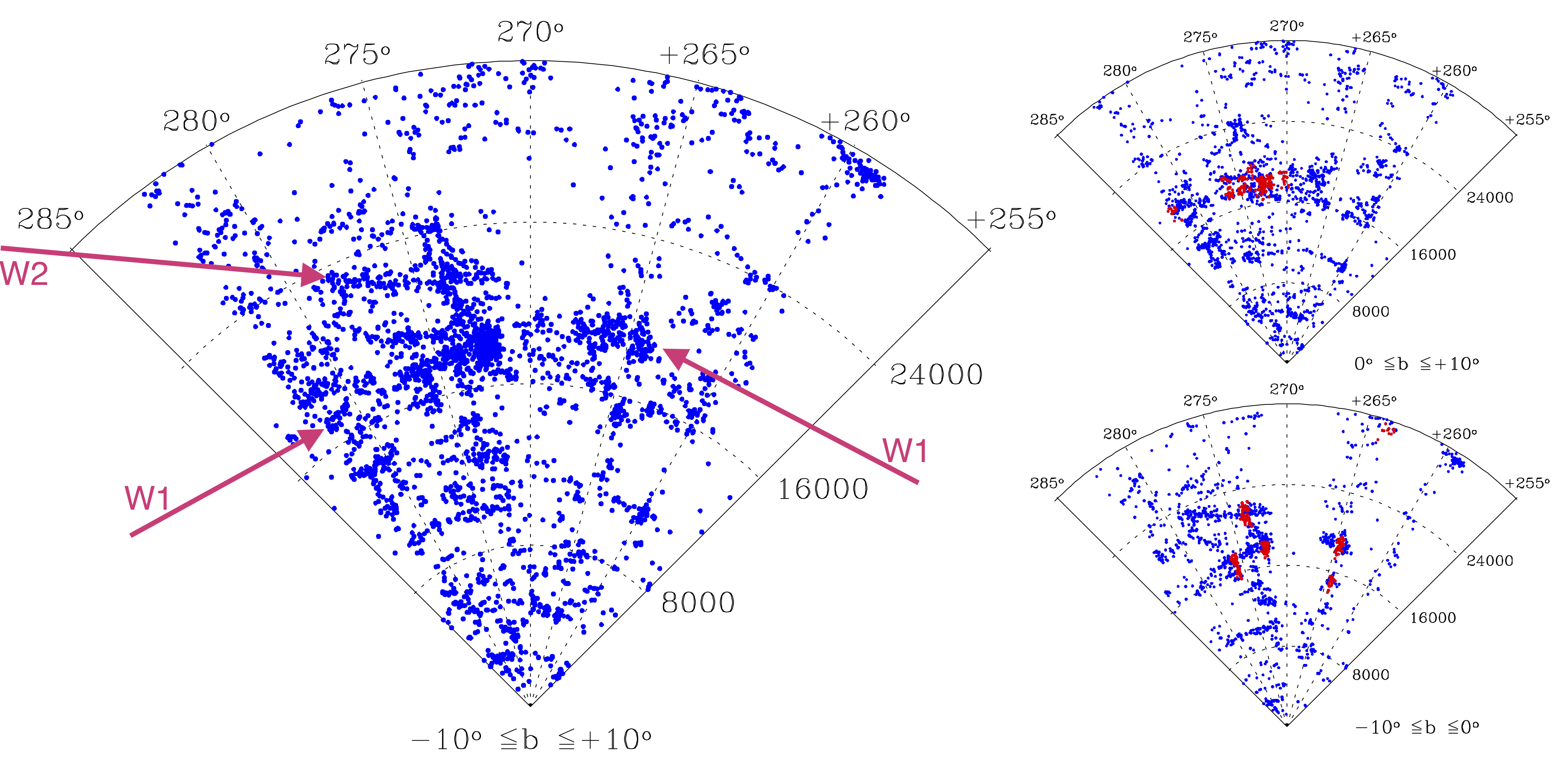}
	\caption{Left-hand panel: redshift cone out to $cz \le 32\,000$\kms\ for the full survey area ($|b| < 10\degr$).
	Right-hand panel: structures above ($b    >10\degr$; top panel) and below ($b<10\degr$; bottom panel) the Galactic Plane; red dots mark galaxies in potential clusters. The primary wall at 18\,000\kms\ is prominent on both sides of the Plane, whereas the higher velocity wall evident below the Plane,  seems to intersect the primary wall in the Plane at $l \le 265\degr$.}
	\label{fig:wedges}
\end{center}
\end{figure*}

Wedge diagrams are presented in Fig.~\ref{fig:wedges} for all the galaxies delimited by the same survey area as Fig.~\ref{fig:vhist}; the right-hand panels show structures above (top) and below (bottom) the Plane. The detail and depth of the large-scale structures unveiled in the ZOA are striking. The overdensity appears to consist of a main broad wall (W1) at $cz \sim 18\,000$\kms\ and a smaller one (W2) at a slightly higher redshift ($cz \sim 22\,000$\kms; see Fig.~\ref{fig:vhist}). The main wall appears both above and below the Plane, while the second wall is present only below the Plane. Most of the gaps in these fairly smooth walls are the result of the sparse sampling to date; see Fig.~\ref{fig:velonsky}. The main wall extends over at least 120~$h_{70}$~Mpc.

Both walls are quite broad, encompassing numerous clusters and groups. The SALT observations confirm seven clusters (five new) within these walls, with indications of further clusters in many of the AAOmega fields, despite undersampling due to fibre positioning constraints. At present the survey does not reach much below the characteristic magnitude at the mean Vela Supercluster (VSCL) distance (see next section). A preliminary clustering analysis based on these redshift data, including an assessment of the velocity dispersion, cluster-centric distribution of redshifts, the spatial concentration, and the steepness of the extinction-corrected $K_s^o$-band luminosity function (LF), has led to the identification of 20 potential clusters with velocity dispersions $\sigma_v > 400~$\kms\ that could be part of the Vela overdensity (galaxies within an Abell radius are marked as red dots in Fig.~\ref{fig:wedges}). 

\section{Vela as supercluster}\label{sect:discuss}

{\sl Supercluster morphology.} The spectroscopic data are consistent with the Vela overdensity being a supercluster: an elongated, prolate, wall-like structure \citep{Einasto11} with embedded clusters.  Examinations of the walls from various perspectives, including photometric redshift data \citep[2MPZ,][]{2MPZ}, suggest that the two walls below the Plane merge in the obscured region, then emanate above the Plane as one wall. This again is a typical morphology for a large supercluster, which can show multi-branching of filaments and walls \citep{Einasto11}. Both the observed cosmic web and simulations of structure formation \citep{Springel05} predict that massive clusters reside at the confluence of such walls. It is an unfortunate coincidence that the current optical data preclude charting the merger of these walls, which occurs in the most obscured part of the ZOA---we are most likely missing the core of the VSCL. 

{\sl Clusters.} An indicator of the mass of a supercluster is the number of its massive clusters. A total of 20 potential new clusters were identified in our data set (see Fig.~\ref{fig:wedges}), in addition to the two CIZA X-ray clusters just outside of our survey region. Surprisingly, \cite{Kocevski06} list no further CIZA clusters within the VSCL walls. Their catalogue is claimed to be fairly complete to Galactic gas column densities of $N_{\rm HI} \lesssim$~3--5~$\times 10^{21}$ cm$^{-2}$ ($A_{\rm K} \sim 0\fm18-0\fm3$; i.e.\ outside the contours in Fig.~\ref{fig:velonsky}). Above this, the low-energy X-ray bands of ROSAT ($0.1-2.4$~keV) are particularly susceptible to absorption by foreground gas---and the Galactic gas flares to higher latitudes (8\degr--10\degr) over a large part of the Vela survey area (Fig.~\ref{fig:velonsky}). To one side of the survey, in a region of diameter  $\sim8\fdg3$ centred on $(l,b) = (263\fdg9,-3\fdg3)$, the X-ray detection of  clusters is extremely difficult due to the bright foreground emission from the Vela supernova remnant.

We explored the ROSAT All-Sky Survey (RASS) database for X-ray emission around the position of the clusters identified in our redshift survey and detected X-ray emission for a few of them, even though the exposure of RASS is very shallow ($200 - 700$~s). The most prominent detection is a dynamically young cluster at $cz = 18\,167$ ~\kms\ with a flux of $F_X = 5.3 (\pm 0.7) \times 10^{-12}$ erg s$^{-1}$ cm$^{-2}$ and an X-ray luminosity of $L_X = 4.6 \times 10^{43}$ erg s$^{-1}$ in the $0.1 - 2.4$~keV band. With this luminosity, and with a complex distribution of emission, it resembles the Virgo cluster of galaxies.

The second brightest X-ray cluster has an X-ray luminosity $L_X = 1.5 (\pm 0.4) \times 10^{43}$ erg s$^{-1}$. Based on the precise X-ray LF determined in the REFLEX II survey \citep{Boeh14}, we can calculate the probability of finding clusters above this luminosity in our survey volume, taken generously as the VSCL region with $N_{HI} \le 3 \times 10^{21}$ cm$^{-2}$ . We find that about 0.8 detections should be expected; thus, the two detections we have are indicative of an overdensity.
 
{\sl Overdensity.} One method to arrive at an estimate of the Vela mass-overdensity, and its potential effect on the local velocity field, is through galaxy counts. To pursue this, we restricted the Vela survey sample to 2MASX galaxies and the survey region to latitude strips of $6\degr < |b| < 10\degr$ (to minimize incompleteness biases) over the longitude range $260\degr < l < 285\degr$ (within these strips, the Galactic extinction does not exceed $A_K =0\fm3$ for $>$99\% of the sources). We refer to this restricted sample henceforth as 2M-Vela. We inspected the apparent $K_s$-band counts and found them to be fairly complete to $K_s = 13\fm8$, though with a slight decline in counts fainter than $13\fm4$. We thus limit the Vela sample to $K_s < 13\fm8$, followed by a further cut to account for a maximum extinction of $A_K =0\fm3$ \cite[assuming][]{Schlafly11}---i.e. $K^o_s < 13\fm5$. This is close to the nominal completeness limit of 2MASX of $K_s < 13\fm5$ away from the Plane. The 2M-Vela sample allows us to estimate a lower limit to the galaxy overdensity. 

For comparison, we defined three 2MASX subsamples with the same magnitude cut-offs. These are the full 2MASX galaxy sample (Full), limited to $|b|>15^\circ$ to avoid low-latitude incompleteness, the South Galactic Cap (SGC, $b<-60^\circ$) and another ZOA sample (ZOA; $110\degr < l < 260\degr$). The latter is restricted to the same latitude strips as 2M-Vela, and excludes the Galactic Bulge; this six times larger ZOA area should be subject to similar incompleteness biases to 2M-Vela. 

\begin{figure}
	\includegraphics[height=0.495\columnwidth]{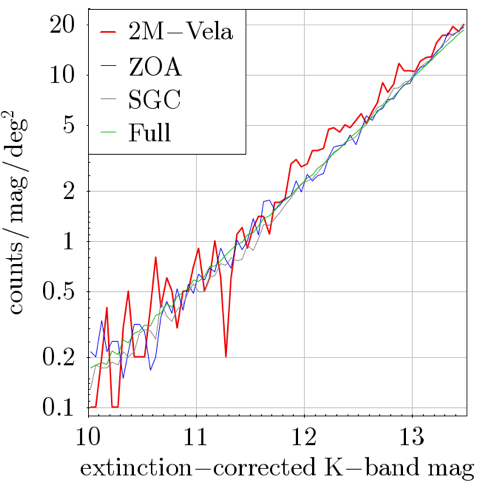} 
    \hspace*{-.1cm}
    \includegraphics[height=0.495\columnwidth]{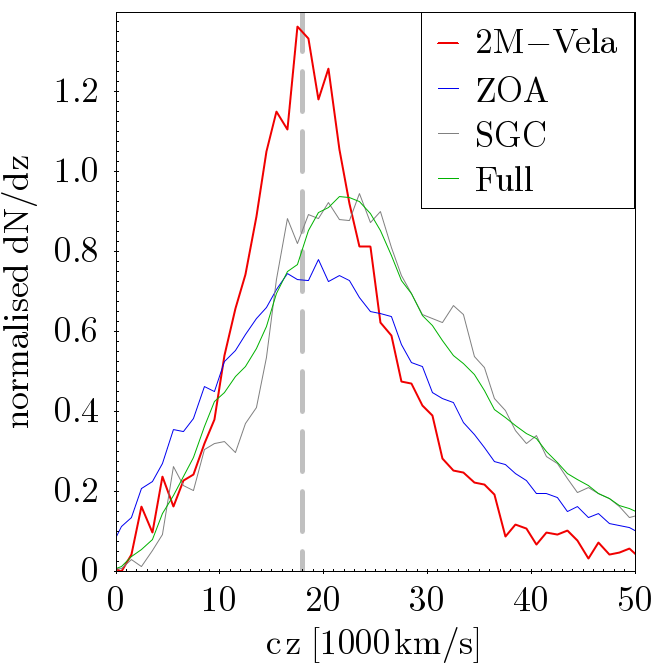}
    \caption{The overdensity of 2M-Vela versus other 2MASX samples. Left-hand panel: direct number counts normalized to the survey areas. The VSCL is notable over the predicted $11\fm8 - 13\fm0$ interval. Right-hand panel:  photometric redshift distributions for the same samples, again normalized by area. The central velocity of the VSCL is indicated by the grey dashed line.} 
    \label{fig:counts}
\end{figure}

We first analysed the number counts per square degree (Fig.~\ref{fig:counts}, left-hand panel). The log$N$ vs $K^o$ relations for all these samples display the expected smooth linear increase up to the completeness limit (even the ZOA sample). However, 2M-Vela reveals, in addition, a clear elevation of counts for $11\fm8 < K^o_s < 13\fm0$ (Fig.~\ref{fig:counts}), exactly where an overdensity at 18\,000\kms\ would reveal its signature. The characteristic magnitude derived by \cite{Kochanek01} from the bright 2MASX catalogue, $M_K^*=-24\fm2$ (adjusted to $h_{70}$), corresponds to $K^o_s < 12\fm8$ for the main VSCL wall (W1). Given the steepness of the LF at the bright end, the increase caused by the supercluster will drop to normal count levels at most one magnitude below $M_K^*$--as observed in Fig.~\ref{fig:counts}. At the faint end the increment will drop more rapidly, exacerbated by increasing incompleteness bias. There is no difference between the enhancement above and below the GP, except for the elevated counts remaining higher at the bright end (for about $\Delta m \lesssim 0\fm2$) below the Plane, the second wall (W2).

We quantify the enhancement by dividing the 2M-Vela counts per deg$^2$ by the respective numbers for the 2MASS comparison samples in the magnitude range in which the VSCL is detectable ($11\fm8 < K^o_s < 13\fm0$). This yields an enhancement in counts by a factor 1.17, 1.21, and 1.25 versus the Full, SGC, and ZOA samples, respectively.

In the second step, we estimate the Vela overdensity $\delta$ ($\delta \equiv \rho/\bar{\rho}-1$)  within a volume shell. We used the 2MPZ catalogue \citep{2MPZ}, which is a combination of spectroscopic redshifts (where available) supplemented by photometric ones. 
Caution in interpreting the data is warranted; large errors in photo-$z$'s ($\sigma_z \sim 0.015$) will smear out structures, making features less prominent and leading to lower overdensity estimates. In addition, the 2M-Vela and ZOA samples will be less complete because of their $K_s^o =13\fm5$ limit. 

The right-hand panel of Fig.~\ref{fig:counts} shows redshift distributions, normalized by survey area, for the discussed samples. The peak at 18\,000\kms\ in the Vela region is clearly visible.
From these data, we determined the density of 2M-Vela and the three 2MASX comparison samples by subdividing them into redshift shells. For a shell of $0.055<z<0.065$, the resulting overdensity is $\delta=0.77$ when compared with the ZOA. As expected the values are slightly lower, $\delta=0.60$ and $\delta=0.51$ compared to 2MASS ($|b| > 15\degr$) or SGC, confirming that the ZOA samples are not as complete as the 2MASS and SGC samples. The values for wider shells are marginally lower (i.e. more diluted).

In summary, the VSCL is significantly overdense in 2MASS galaxy counts, and in a redshift shell centred at its average distance. The overdensity does not vary much on either side of the Plane, giving further substance to the indications that the VSCL extends over at least $25\degr \times 20\degr$ on the sky. 

Despite the limited sampling, we attempt an assessment of how VSCL compares to SSC. Extracting a sample in an area $\sim250$ deg$^2$ around the SSC core in a shell of $0.042<z<0.057$ from the 2MPZ leads to an overdensity of $\delta_\mathrm{SSC}\sim 1.4$. This would, however, decrease 2.3-fold if the SSC were subjected to equivalent selection criteria to Vela (lower 2MASS completeness due to extinction and distance; dilution due to a high fraction of photometric redshifts), i.e. $\delta_\mathrm{SSC}\sim0.6$. We thus conclude that a similarly comprehensive sampling of the VSCL would result in a similar overdensity to SSC, suggesting that the two structures may be comparable, particularly if the central core of VSCL is hidden behind the thickest dust layers.

{\textit{Implications for bulk flows.} The Vela overdensity estimate can be used for a crude assessment of its contribution to the LG velocity. A simple calculation based on linear perturbation theory gives $v_\mathrm{LG} \simeq 50$ \kms. This will likely be a lower limit due to the incomplete sampling. By comparison, the contribution from SSC is estimated -- depending on the method -- to be $v_\mathrm{LG}\simeq 55$ \kms\ \citep{Loeb08} or $v_\mathrm{LG}\simeq 90$ \kms\  \citep{LH11}. The influence of Vela on local bulk flows is thus probably comparable to SSC. A quantitative assessment will, however, require additional observations.

\section{Conclusions}\label{sect:conc}

With close to 4500 new redshifts determined in the ZOA, in Vela, this preliminary data set provides clear evidence for the existence of a very extended ($\sim 115 \times 90~h_{70}$~Mpc) overdensity at a redshift of $\sim$ 18\,000\kms. Its morphology, with a broad main wall and a secondary merging wall, is characteristic of a supercluster. Twenty potential clusters were identified in the walls from our optical spectroscopy. This number constitutes a lower limit, as the surveyed fields cover only $\sim$20\% of the region, while the Galactic foreground gas limits the number of ROSAT X-ray cluster detections. 

The Vela survey galaxy counts reveal an enhancement in counts of $\sim1.2$ compared to three 2MASX comparison samples over the magnitude range of the VSCL, while a redshift shell (based on spectroscopic and photometric redshifts from 2MPZ) shows excess densities of $\delta=0.50-0.77$ (depending on the comparison sample and the width of the shell). The current data set is as yet too sparsely sampled to reliably quantify the cosmological implications of the VSCL. However, preliminary calculations show that an overdensity of this magnitude would reduce the current misalignment of derived clustering dipoles and may play a role in accounting for the observed residual bulk flows \citep[e.g.][]{Scrimgeour16}.

A more comprehensive view of the VSCL requires systematic spectroscopic surveys to bridge the gaps between the current survey fields. The Taipan instrument \citep{Taipan} is ideal for wide-field follow-up spectroscopy, as it can target $\sim$150 objects per $6\degr$ field and will get reasonably complete redshifts, as faint as $r=17\fm5$ with 15\,min integrations. Although the ZOA ($|b| < 10\degr$) does not form part of the Taipan survey, the VSCL, because of its potential science impact, has been selected for observation in the Taipan Science Verification phase in early 2017. 

At higher extinctions, only systematic surveys in the 21\,cm neutral hydrogen line will prevail \citep{HIZOA}. We have proposed using the South African SKA Pathfinder MeerKAT\footnote{\url{http://www.ska.ac.za/meerkat/}} 
in early science mode to survey the most obscured part of the VSCL. Our simulations have shown that this can be realized within reasonable timescales with 32 dishes and does not need to await the full 64-dish array. 

The Vela Supercluster is Terra Incognita, an unknown great continent in the nearby Universe whose outline we are only beginning to discern. We are pursuing an ambitious multi-wavelength program to cover the full width of the ZOA, including the opaque part, in order to uncover the full extent of the Vela Supercluster and determine its implications for cosmology.

\section*{Acknowledgements}

We thank the anonymous referee for insightful comments that improved this paper. 
RCK-K, TJ and MEC acknowledge research support from the NRF. MB is supported through grants \#614.001.451 from the NWO, FP7 \#279396 from the ERC, and \#UMO-2012/07/D/ST9/02785 from the NCN. This publication makes use of data products from 2MASS, which is a joint project of the University of Massachusetts and IPAC/Caltech, funded by NASA and the NSF. The 2MPZ data are available for download at \url{http://ssa.roe.ac.uk//TWOMPZ}. We thank Mark Taylor for the TOPCAT software \citep{TOPCAT}.


\bsp	
\label{lastpage}
\end{document}